\documentclass[showkeys,showpacs,prb,aps,twocolumn,superbib]{revtex4}
\usepackage{graphicx}
\usepackage{color}

\begin{document}
\title{
First-principles study of wurtzite InN$(0001)$ and $(000{\overline 1})$ surfaces
}
\author{Chee Kwan Gan$^1$ and David~J.~Srolovitz$^2$}
\affiliation{
$^1$Institute of High Performance Computing, 1 Science Park Road,
Singapore 117528, Singapore\\
$^2$Department of Physics,
Yeshiva University, New York, NY  10033, USA
}

\date{May 19, 2006}

\begin{abstract}
Density-functional calculations are used to study various plausible
structures of the wurtzite InN $(0001)$ and $(000{\overline 1})$
surfaces.  
These structures include the unreconstructed surfaces,
surfaces with monolayers of In or N, several possible coverages and
locations of In or N adatoms and vacancies. 
The stable structure of the $(0001)$ surface under N-rich conditions
is the unreconstructed, In-terminated, surface, 
while under In-rich conditions the stable surface has a $3/4$ monolayer
of In atoms. 
The stable structure of the InN $(000{\overline 1})$ surface
corresponds to a full monolayer of In atoms in the atop sites (directly
above the N atoms) over the entire range of accessible In (or N) 
chemical potential. 
The atomic structures of the low-energy
structures are also discussed. 
\end{abstract}
\maketitle

\section{Introduction}
Group-III nitrides are of considerable interest for applications in
 laser and light-emitting
diodes\cite{Nakamura96v35,Nakamura97,Taniyasu06v441}.  Alloying
different group-III elements in nitride compounds (e.g., Al$_x$Ga$_{1-x}$N or
In$_x$Ga$_{1-x}$N) provides a means of tuning 
the band gap over a wide spectral range
(e.g., from blue to infrared in the case of In$_x$Ga$_{1-x}$N).
%The importance of these
%materials have led to many theoretical investigations of the
%bulk pure compounds\cite{Wright95v51,Kim96v53,Wright97v82} and 
%their alloys\cite{Ho96v69,Teles00v62,Purton05v15,Gan06_PRB}. The
%effect of defects has also been studied for the pure compounds
%\cite{Gorczyca99v60,Stampfl00v61,VandeWalle04v95} and their
%alloys\cite{Ramos02v14}.  
The growth of high quality group-III nitride compounds can be a considerable
challenge and a barrier to the successful transition of these materials into
applications.  The connection between vapor phase growth 
processes and film quality can usually be traced back to the interplay of growth
conditions and surface structure. While there have been several 
theoretical studies on the surfaces of group-III nitrides, these have largely 
focused upon AlN and
GaN\cite{Northrup96v53,Northrup97v55,Smith97v79,Rapcewicz97v56,%
Fritsch98v57,Wang01v64,Lee03v68,Timon05v72}.  Relatively little 
theoretical analysis has been devoted to the surfaces of InN, largely
because the level of interest in this material has been limited by 
difficulty in producing this material.  

InN is an attractive material for electronic and opto-electronic applications
because it possesses a series of outstanding
properties\cite{Bhuiyan03v94}, such as a small effective mass\cite{Mohammad96v20} (leading to a large electron
mobility and a high carrier saturation velocity).  The steady-state 
drift velocity in InN is $4.2\times 10^7$ cm/s is significantly larger than 
that of many III-V competitors, GaN ($2.9 \times 10^7$~cm/s), AlN 
($1.7\times 10^7$~cm/s), and GaAs 
($1.6\times 10^7$~cm/s)\cite{Foutz99v85,Bhuiyan03v94}.  
Since InN exhibits a high peak overshoot velocity\cite{Foutz99v85}, it is a good
candidate for high-performance heterojunction field-effect
transistors.

The growth of high quality InN is difficult compared to other
III-nitrides, such as AlN and GaN, because it has a relatively  
low dissociation temperature and a high equilibrium
N$_2$ vapor pressure\cite{MacChesney70v5,Ambacher96v14}. 
The lack of suitable
substrate materials also hinders the use of InN as a heteroepitaxial thin film.
Nonetheless, several groups\cite{Davydov02v229,Wu02v80,Nanishi03v42} have
successfully grown good quality InN single crystals.  Interestingly, they have
reported bandgaps of approximately 0.9~eV, which is quite small
compared to the commonly accepted values of 1.8~eV to 2.1~eV.

Since InN does not possess a center of symmetry, 
the (0001) and $(000{\overline 1})$ surfaces are different.  The first is 
In-terminated, while the latter is N-terminated.  Experimental evidence
suggests that polarity (termination) is an important consideration in the growth of 
high quality group-III nitride semiconductors\cite{Saito01v228}.  For example,
surface with opposite polarity exhibit different morphologies\cite{Daudin96v69}.  
Radio frequency molecular 
beam epitaxial (rf-MBE) growth of InN on sapphire\cite{Saito01v228} produces
different polarity surfaces depending on growth temperature (N-terminated at
low temperature, In-terminated at high temperature).  This demonstrates that growth
conditions can modify the relative growth rates of surfaces with opposite polarity. 
The observed change in the relative growth rates of surfaces with opposite polarity 
with temperature
may be associated with different activation energies for growth or may be 
indicative a change in surface structure with growth conditions.
In this work, we determine the equilibrium structure of the important 
(0001) and $(000{\overline 1})$ surfaces of InN as a function of growth conditions.
In particular, we examine the role of In (or N)
chemical potential on the equilibrium surface structure.  We vary chemical potential
rather than temperature, since the range of chemical potentials that are 
experimentally accessible is much larger than that for temperature in 
the most common group-III nitride growth methods (i.e., MOCVD).  

\section{Theoretical models}
The present calculations are performed within the  local
density approximation (LDA) framework, as implemented within the Vienna Simulation
Package (VASP)\cite{Kresse96v54}, employing ultrasoft
pseudopotentials\cite{Vanderbilt90v41}.  The
important\cite{Wright95v51} 4d states for In are included as valence
electrons in the pseudopotentials.  Throughout this work, we employed
a relatively high cutoff energy of 435~eV.  We first optimize the bulk
wurtzite InN structure, yielding the following lattice parameters\cite{Gan06v73}:
 $a=3.518$~\AA, $c = 5.690$~\AA, and
the internal parameter $u=0.379$.  These lattice parameters are in
very good agreement with those found from 
experiment\cite{Davydov02v229}; $a = 3.5365$~\AA\ and $c=
5.7039$~\AA.  

For the surface calculations, our supercells contained six (0001) bilayers
in which the lower four bilayers were fixed in the bulk
configuration.  The upper two bilayers and any adatoms or adlayers
were allowed to relax. Our system is larger than those used by
many others in studies of other group-III nitride surfaces, where typically four
bilayers\cite{Smith97v79,Rapcewicz97v56, Wang01v64,Lee03v68} were 
employed.  To prevent unphysical charge transfer between the top and bottom
slab surfaces, pseudo-hydrogens with fractional charges are used. For the
In-terminated (0001) surface, the dangling bonds in the N layer are
saturated with pseudo-hydrogens with a fractional charge of 3/4.  For
the N-terminated $(000{\overline 1})$ surface, we saturate the
dangling bonds in the In layer with pseudo-hydrogens with a fractional
charge of 5/4.  We employ a $2\times 2 \times 1$ Monkhost-Pack set
sampling scheme for the surface supercell, which gives two irreducible
$k$-points.  The structural optimizations were terminated when the
magnitude of the Hellmann-Feynman force on each ion is less than
1~mRy/Bohr (25.7~meV/\AA).  We perform calculations with vacuum 
thicknesses of both 12.9~\AA\ and 15.7~\AA.  These two sets of calculations 
yield almost identical
results, showing the appropriateness of the vacuum thickness employed.
Hereafter we present the results obtained with a vacuum thickness of
15.7~\AA.

\begin{figure}
\resizebox*{3.0in}{!}{\includegraphics[clip]{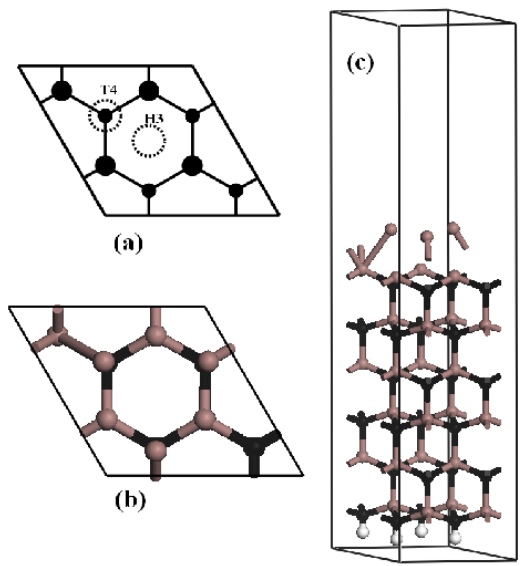}}
\caption{(Color online)
(a) A schematic of the atom positions in the top two atomic planes
of a $2\times 2$ unit cell of the $\{0001\}$ surface.  The
larger and smaller filled circles denote indium atoms on the surface 
and nitrogen atoms one plane down, respectively, for the In-terminated
(0001) surface.  The same schematic describes the $(000{\overline 1})$ 
surface, but with the elemental identity of the atoms switched.  
The locations of the H3 (hollow) and T4 (top)
sites are labeled by dotted circles.
{\bf (b)
A view of an In trimer residing 
on the T4 sites from above.
The gray and black atoms denote the In and N atoms, respectively. Notice that
a nitrogen atom is exposed in the lower right corner of the image.
(c) The side view of the  $2\times 2$
unit cell. Four pseudo-hydrogen atoms (shown in white) are found at the 
bottom of the cell.
}
}
\label{fig:h3t4}
\end{figure}
\begin{figure}
\resizebox*{3.0in}{!}{\includegraphics[clip]{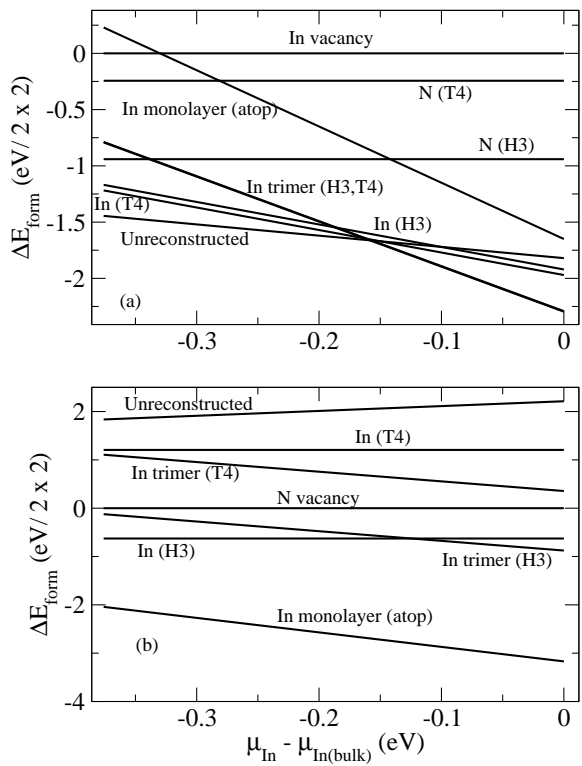}}
\caption{
The relative formation energies for the $2 \times 2$ 
unit cell of Fig.~\ref{fig:h3t4}(a) as a function of the In chemical potential
for the (a) (0001) and (b) $(000{\overline 1})$ surfaces.  
The formation energies
are all referenced to that of the surface with a single vacancy on the
surface atomic plane.
}
\label{fig:1-bar1}
\end{figure}

We consider various vacancy, adatom, trimer, and
monolayer surface structures that satisfy the electron
counting rule\cite{Northrup97v55}. It is useful to refer to Fig.~\ref{fig:h3t4}(a),
which shows two (0001) layers of atoms in a $2 \times 2$ surface unit cell,
in order to understand the structures. 
The surfaces considered include a
perfect surface [as shown by the black circles 
in Fig.~\ref{fig:h3t4}(a)], a monolayer of In (or N)
atoms directly above each In (or N) terminating atom  [i.e., this is 4 atop 
atoms in the $2 \times 2$ cell of Fig.~\ref{fig:h3t4}(a)], a quarter-monolayer of
adatoms with the adatoms directly atop one of the In (or N) atoms (T4 sites)
in the $2 \times 2$ cell, a quarter-monolayer of
adatoms with the adatoms centered between three surface atoms 
(H3 sites), a single vacancy in the terminating plane of the 
$2 \times 2$ cell, and a triangle of three In (or N) atoms on H3 or 
T4 sites (i.e., a trimer). The relaxed structure of a representative case of 
an In trimer on the T4 sites is shown in Figs.~\ref{fig:h3t4}(b) and \ref{fig:h3t4}(c).

The relative formation energies of the different surfaces are calculated
within the thermodynamically allowed range of the In
chemical potential $\mu_{\rm In}$.  
The chemical potentials of In and
N are related through\cite{Qian88v38}
\begin{equation}
\mu_{\rm In} + \mu_{\rm N} = \mu_{\rm InN(bulk)} .
\label{eq:1}
\end{equation}
Operationally, we can define the chemical potential 
$\mu_{\rm InN(bulk)}$ as the total energy per
cation-anion pair $E({\rm InN}_{\rm bulk})$ in  wurtzite InN at 0~K.  The
chemical potential of In is bounded above by that of the pure, tetragonal
In metal\cite{Graham56v84}, denoted by $\mu_{\rm In}^0 = E({\rm In})$.  
Similarly, the chemical potential of N is
bounded above by that of a nitrogen molecule N$_2$, denoted by $\mu_{\rm N}^0 = \frac{1}{2}E({\rm N}_2)$.  
The enthalpy of formation of  InN is simply  
\begin{equation}
 \Delta H = \mu_{\rm InN(bulk)} -
\mu_{\rm N}^0 - \mu_{\rm In}^0 
\label{eq:2}
\end{equation}
where $E({\rm In})$, $E({\rm N_2})$, and $E({\rm InN_{bulk}})$ are determined
from total energy calculations within LDA.
The calculated enthalpy of formation of InN, $\Delta H=-0.376$~eV, agrees reasonably well  with the
experimental value\cite{Harrison89,Kittel96} of $-0.28$~eV.  The fact that
these calculations suggest that InN is more strongly bound than in experiments
is not surprising since the LDA tends to lead to overbinding. 
The small enthalpy of formation implies
that InN has a low thermal stability compared with other group-III
nitrides such as GaN and AlN where the experimental formation
enthalpies\cite{Harrison89,Kittel96} are $-1.23$~eV (i.e., 4 times
larger than for InN) and $-3.21$~eV (i.e., 11 times larger than for InN),
respectively.

The formation energy of any structure is calculated from
\begin{eqnarray}
E_{\rm form} &=& E_{\rm opt} - n_{\rm In}\mu_{\rm In} - n_{\rm N}\mu_{\rm N} \nonumber \\
&=&E_{\rm opt} - \mu_{\rm In}(n_{\rm In}-n_{\rm N})-\mu_{\rm InN(bulk)}n_{\rm N},
\label{eq:3}
\end{eqnarray}
where $E_{\rm opt}$ is the total energy of the structure resulting
from the optimization of the structure with respect to atomic positions
and $n_{\rm In}$ 
and $n_{\rm N}$ are the number of In and N atoms in the structure, respectively.
We have used Eq.~(\ref{eq:1}) to eliminate 
$\mu_{\rm N}$  to arrive at the second line of 
Eq.~(\ref {eq:3}). 
Equations (\ref{eq:1}) and (\ref{eq:2}) show that
the  thermodynamically allowed range of the chemical potential of In is 
$\Delta H + \mu_{\rm In}^0 \le \mu_{\rm In} \le \mu_{\rm In}^0 $.

\section{The $(0001)$ Surface}
The relative formation energies for the $2 \times 2$  InN $(0001)$ 
unit cell of Fig.~\ref{fig:h3t4}(a) are shown in Fig.~\ref{fig:1-bar1}(a) for
the surface structures described above as a function of the In 
chemical potential.  In this figure, we reference the energy
to that of the $2 \times 2$  InN $(0001)$ unit cell with a single In vacancy
in the surface atomic plane. This simplifies
the comparison of energies between the different structures, 
as well as leads to the  cancellation of
the contribution to $E_{\rm opt}$ in
Eq.~(\ref{eq:3}) from the chemical potential of pseudo-hydrogen atoms.
We omit the curves for the structures
of energies greater than 0.25 eV.
Note that the slopes of the lines in Fig.~\ref{fig:1-bar1} simply follow
from the fact that the different structures have 
different numbers of In 
and N atoms (see Eq.~(\ref{eq:3})).

The surface with an N adatom on the H3 site has a lower energy
than that with an N adatom on the T4 site by
0.70~eV/$2\times 2$.  This is largely due to a stronger electrostatic
repulsion between the T4 nitrogen adatom and the nitrogen atom below.
Similar behavior has been observed for the AlN\cite{Northrup97v55} $(0001)$
and GaN\cite{Rapcewicz97v56} $(0001)$ surfaces.
% energy differences of 1.51~eV/$(2\times 2)$ and 0.9028~eV/$(2\times 2)$, 
The surface with an In adatom
on the T4 site has a lower energy than that  with an
In adatom on the H3 site since there is an electrostatic attraction
between the In adatom on the T4 site and the N atom below. The
surface with a full In monolayer with atoms on the atop sites has
a relatively high energy compared with the other (0001) surfaces.

Under N-rich conditions (low In chemical potential), the bare 
$1\times 1$ (In terminated) surface [as shown in Fig.~\ref{fig:h3t4}(a)]
has the lowest energy.
Our calculations also show that surfaces in which the In trimer resides
on either the T4 or H3 sites yield essentially the same energy, 
despite the fact that In atoms are directly above N atoms in 
the T4 In trimer model. Similar behavior is observed for the
AlN$(000{\overline 1})$\cite{Northrup97v55} surface
where Al trimers on the T4 or H3 sites have the same energy.
This shows that arguments based solely on electrostatic interactions 
are not sufficient to explain even the qualitative behavior of 
many of the adatom reconstructed surfaces.
Figure~\ref{fig:1-bar1}(a) shows that under In-rich (large
In chemical potential) conditions, 
the stable (0001) surface have In trimers, located on either 
T4 or H3 sites.  
The present results for the stable (0001) InN surface are 
significantly different than the same surface in other group-III nitrides
across the entire range of chemical potentials.  
For AlN$(0001)$ \cite{Northrup97v55} and
GaN$(0001)$\cite{Rapcewicz97v56} surfaces, the surface with an N
adatom on the H3 site is found to be energetically favorable under
N-rich conditions, while the surface structure with a metal atom
(i.e., Al or Ga) on the T4 site is energetically favorable under
metal-rich conditions.  

Examination of the unreconstructed InN $(0001)$ surface 
[as shown in Fig.~\ref{fig:h3t4}(a)] shows that
the In-N bilayer spacings for the first and second bilayers are
0.654~\AA\ and 0.655~\AA, respectively.  These are slightly smaller
than the ideal, bulk spacing of 0.688~\AA.  The spacing between 
In-In atomic planes are 2.862~\AA\ and 2.825~\AA\ for the first and 
second pairs, as compared with a bulk spacing of 2.845~\AA.  
This shows that the spacing between the first In-In
layer expands, while the second contracts, relative to the bulk spacing.  
Similar oscillatory surface relaxations are observed in other systems 
(e.g., see Ref. \onlinecite{Sun03v68}).

For the InN $(0001)$ surface with a single In vacancy on the surface layer,
we find that the top-layer In atoms undergoes
$sp^2$ bonding configuration by relaxing vertically. This reduces the
bond length between the threefold-coordinated In and the
threefold-coordinated N atoms from the ideal value of 2.145~\AA\ to
2.050~\AA.  This 4.4~\% reduction in length is comparable to that for
the case of AlN\cite{Northrup97v55}, where a 6\% contraction in the 
distance between the Al and N atoms in the vacancy model of AlN$(0001)$
surface was observed.  The vertical movement of the top layer
atoms reduces the top In-N bilayer spacing from the ideal value of
0.688~\AA\ to 0.350~\AA.

The surface for with an In trimer on the T4 or H3 sites
have the same energy and are the most stable surfaces at large
In chemical potential, as discussed above.  
It is interesting to note that the interlayer spacing between the trimer
In atoms and the In atoms in the next plane are 2.61~\AA\  for 
the T4 site trimer and  2.60~\AA\ for the H3 site trimer.  Hence,
these two surface configurations are very similar both structurally
and energetically.

\section{The $(000{\overline 1})$ Surface}

The relative formation energies for the $2 \times 2$  InN $(000{\overline 1})$ 
unit cell of Fig.~\ref{fig:h3t4}(a) are shown in Fig.~\ref{fig:1-bar1}(b) for
the surface structures described above as a function of the In 
chemical potential.  In this figure, we reference the energy
to that of the $2 \times 2$  InN $(000{\overline 1})$  unit cell with a 
single N vacancy in the surface atomic plane.
No results are presented for the surfaces with N adatoms or N trimers
since these systems are unstable.  
Northrup {\it et al.}\cite{Northrup97v55} argued that the
relatively small size of the N atom makes surfaces with N adatoms or
trimers on the AlN$(000{\overline 1})$ highly unstable.  It is
therefore not surprising that our calculations also show that N trimer
or N adatom models are not stable since the single-bond covalent
radius\cite{Cotton95} of In (1.48~\AA) is considerable larger 
than that of Al (1.30~\AA).

The relaxed unreconstructed InN $(000{\overline 1})$ surface
(N-terminated) has a very large  formation energy, relative to most
of the other surface structures.  
We find that the surfaces with an In adatom on the H3 site has an 
energy that is 1.83~eV/$(2\times 2)$ lower than that with the In 
adatom on the T4 site. 
This may be attributed to the fact that there is greater electrostatic 
repulsion between the T4 In adatom and the In atom directly below it
than between the H3 In adatom and the more distant In atoms two
atomic planes below.
Similar argument also holds for In trimer models, where the H3 In
trimer has a lower in energy than the T4 In trimer by
1.23~eV/$(2\times 2)$.  
The surface with an In monolayer with the atoms on
the atop sites has the lowest energy across the entire allowed range of
In chemical potential.  
The next lowest energy structure under N-rich (low In chemical potential)
conditions is surface with a single In adatom in the $2\times2$ unit cell 
on the H3 site.  
These results contrast with the results for other group-III
nitrides $(000{\overline 1})$ surfaces.  In AlN and GaN, the lowest
energy surface structure corresponds to an Al
monolayer (atop sites occupied)\cite{Northrup97v55} or Ga monolayer\cite{Gan06_unpublishedB}  only under metal-rich conditions.  
However, under N-rich conditions, the $(000{\overline 1})$ surfaces 
with  a single Al or Ga atom on the H3 site in the $2\times2$ unit cell have
the lowest energy.

Examination of the InN $(000{\overline 1})$ surface  with an In monolayer 
with the In atoms on the atop sites is the lowest energy structure over the 
entire chemical potential range, as mentioned above.  
The In-N bilayer spacing for the top two bilayers 
are 0.519~\AA\ and 0.510~\AA, as compared with a bulk spacing of
0.668 \AA.  The In-In atomic plane separations are
2.482~\AA, 2.423~\AA, and 2.593~\AA\ for the first three pairs
(counted from the surface), as compared with the bulk value of 2.845~\AA. 
This is a substantial contraction that oscillates as it decays slowly
into the bulk.

The surface structure corresponding to an In adatom in the H3 site
also has a relatively low energy. 
This adatom is situated 0.808~\AA\ above the triangle of three N 
atoms in the plane below.  
The corresponding In-N bond length is 1.980~\AA. 
The three N atoms that form this triangle (i.e., nearest neighbors
of the In H3 adatom) are displaced upward, toward the In adatom
by 0.424~\AA\ relative to the other, threefold-coordinated surface
N atoms. 
 
In the InN$(000{\overline 1})$ surface structure with a single 
N-vacancy in the $2\times2$  unit cell, the  top-layer N atoms 
undergo substantial vertical relaxation.
In this structure, the N-In bilayer spacing is 0.328~\AA\ as compared 
with the bulk spacing of  0.688~\AA. 
The bond length between the threefold-coordinated N and In atoms 
is also reduced from the ideal value of 2.145~\AA\ to 1.898~\AA. 
This is an 11.5\%\ reduction in bond length, which is much larger than
the 6\%\ contraction observed in the N vacancy surface structure 
of AlN$(000{\overline 1})$\cite{Northrup97v55}.  
This large contraction may be attributed to the system trying to move 
toward a configuration that is $sp^2$ bonded.

\section{Conclusions}
We have performed density-functional calculations to 
determine the stable structure of both the  InN $(0001)$ 
and $(000{\overline 1})$ surfaces as a function of the
In (or N) chemical potential. We considered twelve plausible
surface structures.  Several earlier studies considered 
the structure of this surface in other group-III nitrides. 
While structurally similar, InN is quite distinct, as indicated 
by a much lower heat of formation than the others (BN, AlN,
GaN).

For the $(0001)$ surface, we find that the relaxed,
unreconstructed InN $(0001)$ (In terminated) surface 
is stable under N-rich (low In chemical potential) conditions.
On the other hand, under In rich (large In chemical potential)
conditions, the stable structure corresponds to a trimer of
In adatoms on either H3 or T4 sites (this is $3/4$ of an In monolayer).
These stable $(0001)$ surface structures differ from those 
found for two other closely related group-III nitrides
over the entire range of accessible chemical potentials.
The stable structure of the InN $(000{\overline 1})$ surface corresponds
to a monolayer of In atoms directly above the surface N atoms.
This structure is stable over the entire range of accessible chemical
potentials. 
In AlN and GaN, the same surface structure was only found to be
stable under metal-rich conditions.

The present results provide guidance for the exploitation of 
different thin film growth methods/conditions in order to exploit the 
dependence of equilibrium reconstructions on chemical potential.
Unlike in GaN or AlN, the stable structure of the important $(0001)$
surface can be manipulated by changes in growth condition.

\section{acknowledgments}

The authors gratefully acknowledge useful discussions with
Y.~Y.~Sun on surface calculations.
This work was supported by Visiting Investigator Program, Agency for
Science, Technology and Research (A*STAR), Singapore.

%\bibliographystyle{apsrev}
%\bibliography{phase}

\end{document}